\documentclass[prd,nofootinbib,superscriptaddress,12pt]{revtex4}

\usepackage{setspace}
\usepackage{graphicx}
\usepackage{dcolumn}
\usepackage{amsmath}
\usepackage{epsfig}
\usepackage{amsfonts}
\usepackage{amssymb}
\usepackage{revsymb}
\usepackage{verbatim}

\usepackage[T1]{fontenc} 
\usepackage[utf8]{inputenc} 
\usepackage{lmodern}

\usepackage[T1]{fontenc} 

\usepackage{ae} 

\newcommand{\mnras}{Mon. Not. Royal Astron. Soc. }
\newcommand{\aap}{Astron. \& Astrophys. }

\newcommand{\apjl}{Astrophys. J. Letters }

\begin{document}


\title{ Mediatrix method for filamentation of objects in images\\
}



\author{C. R. Bom}
\email{debom@cbpf.br}
\affiliation{Centro Federal de Educa\c{c}\~ao Tecnol\'ogica Celso Suckow da Fonseca \\ Rodovia M\'ario Covas,
lote J2, quadra J - 23810-000 Distrito Industrial de Itagua\'{\i}, Itagua\'{\i}, RJ}
\affiliation{Centro Brasileiro de Pesquisas F\'\i sicas, Rua Dr. Xavier Sigaud 150, Rio de Janeiro, RJ - 22290-180, Brasil}

\author{M. Makler}
\email{martin@cbpf.br}
\affiliation{Centro Brasileiro de Pesquisas F\'\i sicas, Rua Dr. Xavier Sigaud 150, Rio de Janeiro, RJ - 22290-180, Brasil}

\author{Marcelo P. de Albuquerque}
\email{marcelo@cbpf.br}
\affiliation{Centro Brasileiro de Pesquisas F\'\i sicas, Rua Dr. Xavier Sigaud 150, Rio de Janeiro, RJ - 22290-180, Brasil}




\begin{abstract}

We describe the Mediatrix filamentation method, an iterative procedure that decomposes 
image shapes in filaments 
over their intensity ridgeline along their main direction 
using perpendicular bisectors.
From this decomposition several morphological features can be derived, such as the 
length along the main direction 
and thickness of the object and, for curved objects, 
estimates of its center of curvature and curvature radius.
As an example, we apply this technique to
arc-shaped objects: simulated gravitational arcs. 


\end{abstract}

\maketitle

\section{INTRODUCTION}

A reliable and robust set of characteristic features for object analysis is a recurrent issue in image processing.
For example, the definition of basic morphological features, such as the length $L$ and width $W$ of an object, is useful in many applications and may have multiple definitions, which may be useful in different contexts. In the case of curved shapes, one may also define 
a curvature 
center and radius. 
In this work we describe a 
technique to decompose shapes (images of objects) into a set of filaments to extract features such as the length $L$ and width $W$ and determine a curvature center. The method is well suited for curved shapes, though it is applicable to more generic shapes.%

A good example of such curved shapes in an astronomical setting is given by gravitational arcs. These objects are a particular case of the Strong Lensing effect~\cite{Schneider,Mollerach,SThe}, which occurs when a distant source in the Universe (typically a galaxy), an intervening mass distribution (a galaxy or galaxy cluster) and the observer are closely aligned. This matter distribution curves the space-time, acting as a lens.
Gravitational arcs are useful tools to unveil the matter distribution of galaxies and galaxy clusters (e.g.~\cite{coe,carrasco}) and the cosmological model (e.g. ~\cite{jullo,mene,yamamoto}).
Since they are very rare objects, several imaging surveys have been carried out on the celestial sphere to find such gravitational arc systems (e.g., \cite{Nord:2015uju,SOGRAS,Bayliss2012,more11,sdss1, Smith2005, Zaritsky03}) and also in an automated fashion \cite{more11,Horesh2010,seidel,alard,lenzen}. The development of tools to extract features for arc-like objects, such as those derived from the method presented here, is also an important step in developing arcfinders \cite{AFcompMGM, Bom:2016lqe}.

In this paper we describe the Mediatrix Filamentation method (Section 2) and a few morphological features derived from it (Section 3). We present a concrete example of application on the images of three simulated arcs (Section 4). Finally we discuss possible generalisations and other applications of the method (section 5).

\section{MEDIATRIX FILAMENTATION METHOD \label{method}}


The method consists in decomposing objects into a set of line segments that follow its ridgeline, which may be used to determine the object ``spine''. For the purposes of this work we will consider pixelated objects, though the method may be applied to a more general object, not necessarily to a digital image.
The only strong requirement is that the object should have a well defined boundary, i.e., it must be separated from the background.

The Mediatrix method was designed to characterize arc-like objects. Its key idea is inspired on a basic geometrical property of the perpendicular bisector of pairs of points on a circle, namely that these lines, for any set of pairs of points, intersect at the circle center. Therefore if an elongated object can be decomposed into a set of points along its longer direction, and if this object has a shape close to an arc segment, the perpendicular bisectors of pairs of these points should intersect close to the center of curvature.

The procedure to filament the object is to recursively obtain the perpendicular bisector of pairs of points on the image. For two points $P_1=(x_1,y_1)$ and $P_2=(x_2,y_2)$, the perpendicular bisector is a straight line $y = mx + b$ perpendicular to the segment $\overline{P_1 P_2}$, intersecting this segment in its mid point and whose coefficients are given by:

\begin{equation}
\label{coeficiente} m= - \frac{ x_2-x_1}{y_2-y_1},
\end{equation}

\begin{equation}
\label{coeficiente2} b=   \frac{ (y_1 + y_2)-m(x_1+x_2)}{2}. 
\end{equation}
\\

The Mediatrix method works in $n$ iteration steps. Each step is a new Mediatrix level. In the following, we describe the first few levels as an example.

\begin{figure}[htb]
       \includegraphics[width=13.5cm]{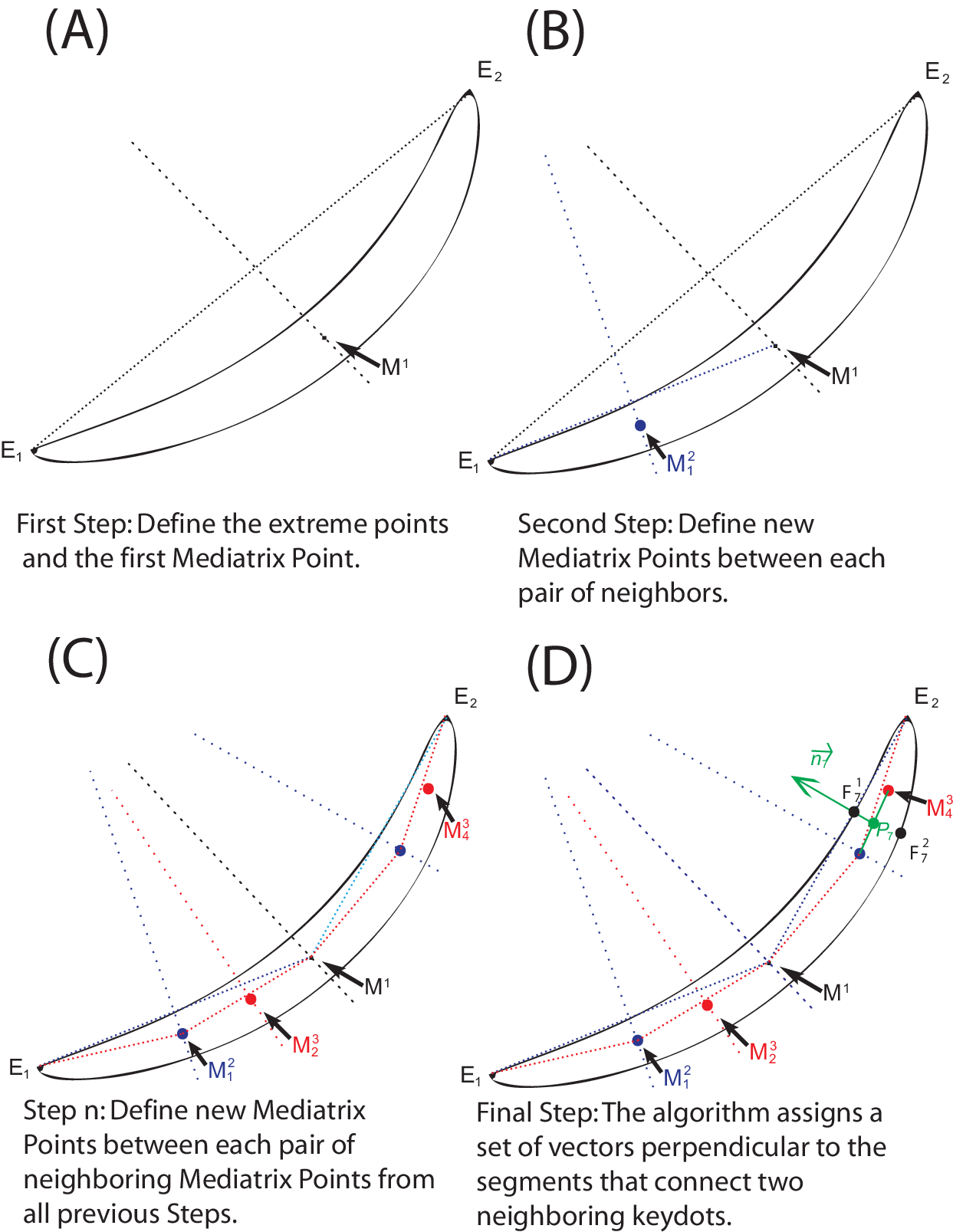}
       \caption{
Steps in the Mediatrix Filamentation method. After $n$ iterations, the method determines a set of $2^n$ points defined by the maximum of intensity along the $2^n$ perpendicular bisectors and $2^n$ vectors perpendicular to neighbouring points with magnitude given by the distance between these points. For clarity, only some points are shown on the figure, which illustrates the steps for $n=3$.}
       \label{Step1}
   \end{figure}

In the first step we determine the extreme points, $E_1$ and $E_2$ of the object (i.e. the two points most distant from each other). 
Several methods have been considered to determine the extreme points of an object (see, e.g., \cite{MaxDistTest}). Here we use the ``farthest-of-farthest'' method, by which $E_1$ is defined as the most distant point from a reference point on the object (e.g., the brightest pixel on the image or its geometrical center), whereas $E_2$ is defined as the pixel on the object farthest from $E_1$. After that, we evaluate the first perpendicular bisector between these two points. The first \textit{Mediatrix point} $M^1$ is defined as the brightest pixel of the object along the perpendicular bisector. Since the object is a set of pixels, we take the brightest pixel located at a distance $d \leq \alpha \Delta p$ from the perpendicular bisector, where $\Delta p$ is the pixel scale and $\alpha$ is chosen as $\alpha =\frac{\sqrt{2}}{2}$. The first Mediatrix Point $M^1$ is shown in Fig. \ref{Step1}(A) for an arc-shaped object (more specifically, an ArcEllipse \cite{arcfitting}).

In the second step the perpendicular bisector is now calculated with respect to $E_1$, $M^1$ and $M^1$, $E_2$. These two perpendicular bisectors now define the next two Mediatrix Points: $M^{2}_{1}$ and $M^2_2$ using the same criteria we used to define $M^1$ (Fig 1B). The upper index refers to the iteration level and the lower index is a label to identify the points. Moving forward to the third step, presented in Fig 1(C), we start from the previous set of Mediatrix Points $M^1$, $M^{2}_{1}$, $M^2_2$ and the two extrema $E_1$ and $E_2$. Those points are used to define new Mediatrix Points in the third level $M^3_i$ are calculated by picking the highest intensity pixel close to the perpendicular bisector between two neighboring points of the previous iteration level. The algorithm continues defining new Mediatrix Points $M^j_i$, corresponding to the $i$-est point in the $j$-est iteration level, in higher iteration levels until reaching a specified final step $n$. In Fig 1(D), we present the last step for $n=3$ (as in the previous panels some points were omitted not to crowd the figure). The collection of Mediatrix Points together with the two extrema are then named {\it keydots}. From the keydots, the object is decomposed into $N=2^n$ segments or filaments. Each segment connects a keydot to its neighbors. The algorithm outputs a set of vectors $\vec{n}_j$, where $j$ varies from $1$ to $N$. Those vectors are perpendicular to the segment that connects a keydot to its neighbor with origin in the midpoint of this segment and norm equal to the length of the segment. This is shown in Fig. 1(D) for $\vec{n}_7$, where $|\vec{n}_7|$=$|\overline{M^3_4 M^2_2}|$.


\section{FEATURES OBTAINED FROM THE MEDIATRIX METHOD \label{measurements}}

Using the outputs of
the Mediatrix filamentation method, the object length, $L$, is defined as:
\begin{equation}
\label{length} L= \sum^N_{j=1} |\vec{n}_j|.
\end{equation} 

The points $\vec{F}^1_j$ and $\vec{F}^2_j$, represented in Fig. 1(D) for $j=7$, are defined as the two
most distant points from the set of pixels
along the perpendicular bisector associated to $\vec{n}_j$. Using those points, it is possible to estimate the average thickness $W$:

\begin{equation}
\label{thickness} W= \frac{1}{N} \sum^N_{j=1} |\vec{F}^1_j-\vec{F}^2_j|.
\end{equation}

For arcs constructed from circle segments (as in the case of the ArcEllipse shape \cite{arcfitting}), all perpendicular bisectors intercept at the center of curvature. 
Therefore, we may define as the object center of curvature the 
average of the points $C_{ik}=(x_{ik},y_{ik})$ defined by the intersection of the lines in the $\vec{n}_i$ and $\vec{n}_k$ directions. The center coodinates $C_a=(x_a,y_a)$ are thus given by:
\begin{equation}
\label{Media1} x_a= \frac{(N-2)!}{N!} \sum^N_{i=1}\sum^N_{k\neq i} x_{ik}
\end{equation}

\begin{equation}
\label{Media2} y_a= \frac{(N-2)!}{N!} \sum^N_{i=1}\sum^N_{k\neq i} y_{ik}
\end{equation}

An alternative way to define a curvature center is by using the median intercept point $C_m=(x_m,y_m)$, where $x_m$ is defined as the median of the values of $x_{ik}$ and $y_m$ is given by the median of $y_{ik}$. 

From these two ``center of curvature'' definitions we may assign a
curvature radius $R_a$ and $R_m$, associated to $C_m$ and $C_a$, as the distance from the first Mediatrix filamentation point $M^1$ to $C_m$ or $C_a$ respectively, i.e., 
$R_a= |\overline{M^1 C_a}| $ and $R_m= |\overline{M^1 C_m}| $.

\section{EXAMPLE OF APPLICATION: SIMULATED GRAVITATIONAL ARCS \label{app}}

\begin{figure}[!htb]
       \includegraphics[width=7cm]{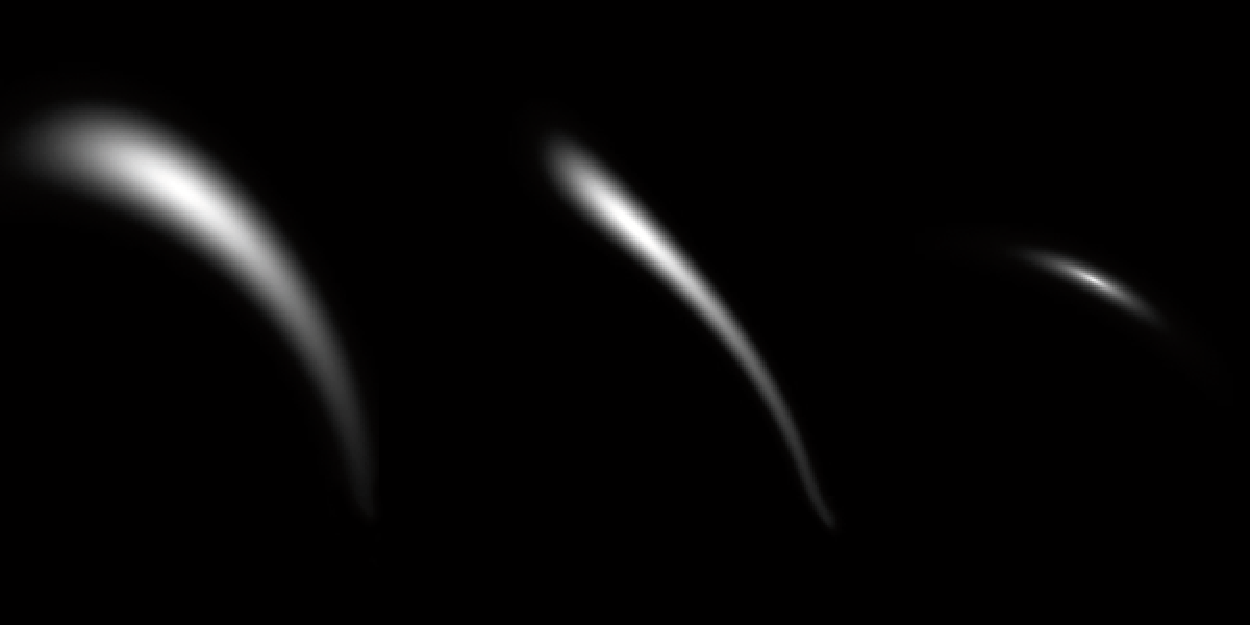}
       \caption{
       Arcs A, B, and C respectively used as input for the Mediatrix method in the examples of this paper.}
       \label{ex02}
   \end{figure}

As an example of application of the Mediatrix filamentation method, we consider 3 arc-shaped objects produced through the gravitational lensing effect (see Fig. \ref{ex02}).
These arcs were generated using the \textit{AddArcs}~\cite{addarcs} pipeline to simulate gravitational arcs.

 \textit{AddArcs} is a code that simulates the surface brightness distribution of  gravitational arcs, e.g., high elongated lensed objects, using the galaxy cluster abundance (i.e. the lenses) provided by cosmological simulations and background galaxies (the sources) with morphological parameters and redshift distribution obtained from the Hubble Ultra Deep Field Survey \cite{UDFBeckwith,UDFCoe}. The source brightness distribution is modelled by S\'{e}rsic profiles~\cite{sersic} (which in principle extend to infinity) with elliptical isophotes. The lens model is given by a Navarro-Frenk-White density profile (NFW, \cite{nfw1,nfw2}), with elliptical projected density (see, e.g., \cite{caminha}). Given the input models 
for the source and the lens, \textit{AddArcs} controls the software {\it gravlens} \cite{gravlens}
to perform the gravitational lensing calculations, resulting in one or more images of the source.
The identification of individual images and their segmentation 
(i.e. the determination of the image pixels) is carried out using the {\it SExtractor} \cite{sextractor} software, in a similar way as to the one described in \cite{arcfitting}.

\begin{figure}[!htb]
       \includegraphics[width=7cm]{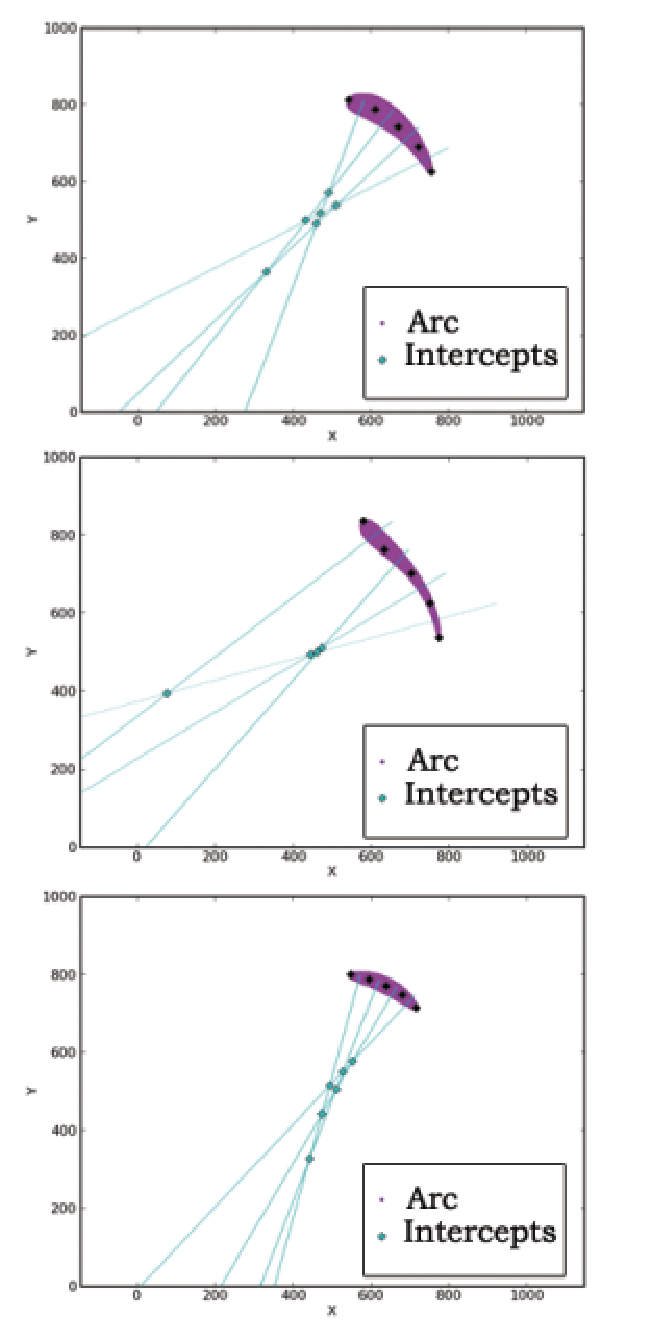}
       \caption{
Keydots (black lozenges), lines defined by the vectors $\vec{n}_j$ (cyan lines), and their intercepts (cyan lozenge), for $n=2$. From top to bottom: arcs A, B and C.}
       \label{ex03}
   \end{figure}
 
In  Fig. \ref{ex03} we show the results of the Mediatrix method for $n=2$ on the 3 example objects mentioned above. The keydots are shown as black lozenge on the objects. We also show the lines spanned by the vectors $\vec{n}_i$ --- perpendicular to the lines connecting neighbouring keydots --- and their intersections, the points $(x_{ik},y_{ik})$, as cyan lozenge.
In order to avoid lines that do not intercept or are almost parallel to each other, the algorithm ignores combinations of $\vec{n}_i$ and $\vec{n}_k$ that are closely aligned, more specifically, we discard pairs with $|\tan\theta_i-\tan\theta_k|\leq 10^{-3}$, where $\theta_i$ and $\theta_k$ are the angles determined by $\vec{n}_i$ and $\vec{n}_k$ with respect to the $x$ axis. From $(x_{ik},y_{ik})$ we define the mean and median center of curvature, $C_a$ and $C_m$, respectively, as described in the preceding section.

The center of the mass distribution that generated the lensing effect, the lens center $C_l=(x_l,y_l)$ is given by \textit{AddArcs}. 
We define the arc radius $R$ as the distance from the first Mediatrix filamentation point $M^1$ to $C_l$, i.e., $R= |\overline{M^1 C_l}| $. 
From gravitational lensing theory it is known that the curvature center of the arcs and the center of mass do not need to coincide, however, for models as the ones used in \textit{AddArcs} (elliptical lenses), they are usually close.
Therefore, we expect the fractional differences $\Delta R_m/R$ and $\Delta R_a/R$, where $\Delta R_m$ ($\Delta R_a$) is the distance between $C_l$ and $C_m$ ($C_a$), to be small for the example objects in consideration.

In table \ref{table1} we show the results for some of the Mediatrix based measurements on the 3 example arcs as a function of $N=2^n$ for up to 5 iterations of the Mediatrix filamentation. Although the results are not statistically meaningful, the behavior is representative of what we see for a large number of simulated arcs. The determination of both $L$ and $W$ is robust, in the sense that these quantities are weakly dependent on $n$ and $n=2$ already provides good estimates for them. As expected $L$ increases with $n$, while $W$ should be more stable, such that $L/W$ has a (weak) increase with $n$.\footnote{This is what we expect for images without noise and where the object $L/W$ is large enough so that many iterations can be done without loss of information.}

On the other hand the determination of the curvature center is not robust and has a strong variation with $n$, having an oscillatory behavior that does not seem to converge. However, the fractional difference between the lens and the arc curvature centers is minimised for $n =2$ and $ n =3$. For these values of $n$, the estimates using the median are better than those using the mean. We defer to another paper a more complete and systematic analysis of the morphological properties of the objects for various shapes and iteration levels.



\begin{table}


\centering

\begin{tabular}{ccccc} 
\hline 

$N$ & $\frac{\Delta R_m}{R}$ & $\frac{\Delta R_a}{R}$ & $L$(pixels) & $L/W$ \\

\hline
\hline
\multicolumn{5}{l}{Arc A} \\

\hline
4  & 0.036 & 0.062& 294.7& 5.6\\ 
8  & 0.153 & 0.126& 295.3& 5.6 \\
16 & 0.144 & 0.092 & 295.7 & 5.6\\ 
32 & 1.09 & 0.230 & 296.8 & 5.7 \\ 

\hline
\multicolumn{5}{l}{Arc B} \\
\hline
4 & 0.100 & 0.096 & 362.4 & 12.6 \\
8 & 0.211 & 0.68 & 363.1 & 12.7 \\
16 & 0.166 & 0.170 & 363.3 & 12.7 \\
32 & 0.307 & 0.168 & 363.9 & 12.7 \\

\hline
\multicolumn{5}{l}{Arc C} \\
\hline
4 & 0.101 & 0.097 & 192.9 & 5.5 \\
8 & 0.051 & 0.093 & 193.1 & 5.5 \\
16 & 0.085 & 0.126  & 193.5 & 5.6\\
32 & 0.198 & 0.209 & 194.7 & 5.6 \\

\hline

\end{tabular}
\caption{Results of Mediatrix measurements for 3 simulated arcs A, B and C. The first column is the number of segments the arc was divided into. The second and third are the ratio between the distance from the lens center to the center of curvature --- calculated using the average $\Delta R_a$ and median $\Delta R_m$ respectively --- and the curvature radius $R$. The fourth column is the arc length and the fifth is the $L/W$ ratio. 
}
\label{table1}
\end{table}

\section{ CONCLUDING REMARKS \label{discuss}}

The Mediatrix filamentation method is a technique to decompose elongated shapes into filaments in the main object direction, defining a ``spine'', which allows one to define a number of morphological features, including the length, width, and curvature center and radius.
%
It may be used in the detection of shapes such as gravitational arcs and to provide a more generic and robust characterization of these objects.%
\footnote{Earlier definitions of the curvature center 
for gravitational arcs
used only 3 object points to define a circle~\cite{arc_catalog}. 
The Mediatrix method allows for definitions of curvature center that use information from many object pixels.}
For example, in ref. \cite{arcfitting} the Mediatrix output is used to set some parameters in 
a method that fits 
the brightness distribution of gravitational arcs using
analytical templates to derive their structural parameters.
Another application is to use parameters derived from the Mediatrix decomposition as inputs for a neural network trained to find gravitational arcs \cite{AFcompMGM, Bom:2016lqe}. 

In the examples of this paper, we considered elongated objects with a well defined ``main direction'', a surface brightness distribution with a clear ``spine'', and noise-free images. However, the method can be extended to more general situations. For example, instead of using the object extrema as 
the first Mediatrix step, other definitions for the first step can be used for objects without a clear single preferred direction,
such as objects with several tips. 
In this case, each pair of tips will define a different set of Mediatrix filaments and the final filamentation can be constructed as the combination of them. For binary images, where the pixels have all the same value, the midpoint of the pixels along the perpendicular bisector (i.e. the midpoint of $F^1_j$ and $F^2_j$ in figure 1) can be chosen as Mediatrix points, instead of the brightest pixel along this line. 

In situations where the fluctuations on the pixel values are significant --- which is often the case in astronomical images, in particular of gravitational arcs --- the determination of the brightest pixel will strongly depend on the  noise. In this case it is necessary to first smooth the brightness distribution using either using Gaussian isotropic kernels \cite{sextractor} or more sophisticated schemes, such as anisotropic smoothing \cite{lenzen}.
In addition, in practice the filamentation procedure cannot be applied to arbitrarily large orders $n$. In particular, we expect to loose directional information when each segment of size $|\vec{n}_i|$ is of the order of the width $|\vec{F}^1_j-\vec{F}^2_j|$. The criteria to stop iterating will also depend on the noise level and on the smoothing scale. A global value of $n$ may be set or otherwise the number of iterations may be determined adaptively, at the level of each segment after a given Mediatrix step (thus choosing if that segment will be further decomposed by a new perpendicular bisector or not).

A more complete and systematic analysis will be carried out in a separate work, using a large number of real and simulated arcs (with noise, background, etc.) to determine the optimal settings for convergence and robustness of the results \cite{Bom2016}. Extensions to other shapes will also be further investigated. However, the main aim of this paper is to present the Mediatrix filamentation method, describing its basic procedures and a few derived morphological quantities, showing an example of application to simple shapes derived by simulating the strong gravitational lensing effect leading to arcs.

\bigskip
\bigskip

{\bf ACKNOWLEDGMENTS:}
\bigskip
\bigskip

C.~R.~Bom would like to thank CNPq for the financial support. M.~Makler is partially supported by CNPq 
(grant  312353/2015-4) and FAPERJ (grant E-26/110.516/2012 and E-26/210.725/2015).
We thank Bruno Rosseto for reminding us of the geometrical proprieties of the perpendicular bisector in a circle and Anupreeta More and Bruno Moraes for useful suggestions to the manuscript. 

\end{document}